\documentclass{emulateapj}

\usepackage{graphicx}
\usepackage{textcomp}

\def\HI{H\,\textsc{i}}

\def\kms{${\rm km~s}^{-1}$}

\begin{document}

\title{Molecular Emission from a Galaxy Associated with a $z \sim 2.2$ Damped Lyman-Alpha Absorber}

\shorttitle{Molecular Emission from a Galaxy Associated with a $z \sim 2.2$ DLA}
\shortauthors{Neeleman et al.}

\author{Marcel Neeleman\altaffilmark{1,2}, Nissim Kanekar\altaffilmark{3}, J. Xavier Prochaska\altaffilmark{1},
Lise Christensen\altaffilmark{4}, Miroslava Dessauges-Zavadsky\altaffilmark{5}, Johan P.U. Fynbo\altaffilmark{6},
Palle M{\o}ller\altaffilmark{7}, Martin A. Zwaan\altaffilmark{7}}
\altaffiltext{1}{Department of Astronomy \& Astrophysics, UCO/Lick Observatory, 
University of California, 1156 High Street, Santa Cruz, CA 95064, USA}
\altaffiltext{2}{Max-Planck-Institut f\"{u}r Astronomie, K\"{o}nigstuhl 17, D-69117, Heidelberg, Germany}
\altaffiltext{3}{Swarnajayanti Fellow; National Centre for Radio Astrophysics, Tata Institute of 
Fundamental Research, Pune 411007, India}
\altaffiltext{4}{Dark Cosmology Centre, Niels Bohr Institute, Copenhagen University, 
Juliane Maries Vej 30, DK-2100 Copenhagen O, Denmark}
\altaffiltext{5}{Observatoire de Gen\`{e}ve, Universit\'{e} de Gen\`{e}ve, 51 Ch. des Maillettes, 
1290 Versoix, Switzerland}
\altaffiltext{6}{Cosmic Dawn Center, Niels Bohr Institute, Copenhagen University, 
Juliane Maries Vej 30, DK-2100 Copenhagen O, Denmark}
\altaffiltext{7}{European Southern Observatory, Karl-Schwarzschildstra{\ss}e 2, D-85748 
Garching bei M\"{u}nchen, Germany}
\email{Email: neeleman@mpia-hd.mpg.de}

\begin{abstract} 
Using the Atacama Large Millimeter/sub-millimeter Array, we have detected CO(3$-$2) line  
and far-infrared continuum emission from a galaxy associated with a high-metallicity ([M/H] = $-0.27$) 
damped Ly-$\alpha$ absorber (DLA) at $z_{\rm DLA}=2.19289$. The galaxy is located $3.5''$ away from the 
quasar sightline, corresponding to a large impact parameter of 30~kpc at the DLA redshift. We use archival 
Very Large Telescope-SINFONI data to detect H$\alpha$ emission from the associated galaxy, and find that the 
object is dusty, with a dust-corrected star formation rate of $110^{+60}_{-30}$~M$_\odot$~yr$^{-1}$. The galaxy's
molecular mass is large, M$_{\rm mol}$ = $(1.4 \pm 0.2) \times 10^{11} \times (\alpha_{\text{CO}}/4.3) 
\times (0.57/r_{31})$~M$_\odot$, supporting the hypothesis that high-metallicity DLAs arise predominantly 
near massive galaxies. The excellent agreement in redshift between the CO(3$-$2) line emission and low-ion 
metal absorption ($\sim 40$~\kms)  disfavors scenarios whereby the gas probed by the DLA shows bulk motion 
around the galaxy. We use Giant Metrewave Radio Telescope H\,\textsc{i}~21\,cm absorption spectroscopy to 
find that the \HI\ along the DLA sightline must be warm, with a stringent lower limit on the spin temperature of 
$T_{\rm s} > 1895 \times (f/0.93)$~K. The detection of C\,\textsc{i} absorption in the DLA, however, also indicates 
the presence of cold neutral gas. To reconcile these results requires that the cold components in the DLA contribute 
little to the \HI\ column density, yet contain roughly 50\,\% of the metals of the absorber, underlining the complex 
multi-phase nature of the gas surrounding high-$z$ galaxies.
\end{abstract}


\section{Introduction}
\label{sec:intro}

Absorption spectroscopy of quasars, to this day, remains the most efficient way to study \HI\
at high redshifts. The strongest \HI\ absorbers, the so-called damped Ly-$\alpha$ 
absorbers \citep[DLAs; e.g.,][]{Wolfe2005} have \HI\ column densities $\geq 2 \times 10^{20}$~cm$^{-2}$
and contain the vast majority ($>$ 80~\%) of \HI\ at high redshifts \citep[e.g.,][]{Peroux2003}.
DLAs are expected to be closely associated with galaxies because of the well-established correlation 
between \HI\ and star formation \citep[e.g.][]{Schmidt1959}, and because in the local universe \HI\ column 
densities comparable to DLAs occur primarily in galaxy disks. However, the nature of the galaxies associated 
with DLAs remains an open question, with simulations \citep[e.g.,][]{Bird2014} predicting dark matter 
halo masses significantly smaller than those inferred from DLA cross-correlation observations 
\citep{Perez-Rafols2018}. 

Over the last three decades, many attempts have been made to image the galaxies associated with DLAs 
directly in emission \citep[e.g.,][]{Lebrun1997,Moller2002,Chen2005}. These studies
have been moderately successful at low redshift, but despite valiant efforts using innovative
observation techniques \citep[e.g.][]{Kulkarni2006,Fynbo2010,Fumagalli2015,Johnson-Groh2016}, fewer than 20 
absorber-galaxy pairs are known at $z \gtrsim 2$ \citep{Krogager2017}. 
With the commissioning of efficient integral field unit (IFU) spectrographs on 10\,m-class telescopes,
the ability to detect galaxy-absorber pairs has significantly improved 
\citep[e.g.,][]{Peroux2011,Bouche2013,Jorgenson2014,Fumagalli2017}. However, all of these studies aim at detecting
the emission from the stars and the ionized regions of galaxies, which may be highly obscured in a dusty
galaxy. The Atacama Large Millimeter/sub-millimeter Array (ALMA) provides a complementary approach,
whereby we can search for longer wavelength emission (e.g., from CO, [C\,\textsc{ii}] lines, and radio continuum),
which is less affected by the presence of dust and arises predominantly from the molecular and atomic gas 
inside the galaxy.

In \citet{Neeleman2016b}, we presented the first detection of molecular emission from a galaxy 
associated with a Ly-$\alpha$ absorber. In subsequent work, we found that molecular emission 
is detected in a large fraction of galaxies associated with high-metallicity absorbers at $z \approx 0.7$, 
and that their gas fraction is significantly higher than that of emission-selected galaxies at these redshifts
\citep{Moller2018,KanekarPrep}. Encouraged by these results, we have targeted three DLAs at $z \sim 2$
with ALMA to search for CO emission from galaxies associated with the absorbers. One of them 
---the DLA toward QSO~B1228$-$113 \citep{Ellison2001}---
is the focus of this Letter; the remaining two systems will be discussed in a future paper. 
Throughout this Letter we assume a standard  flat Lambda Cold Dark Matter cosmology 
with $\Omega_\Lambda = 0.7$, and $H_0 = 70~{\rm km~s^{-1}}$.

\begin{figure}[t!]
\hspace{-0.5cm}
\includegraphics[width=0.5\textwidth]{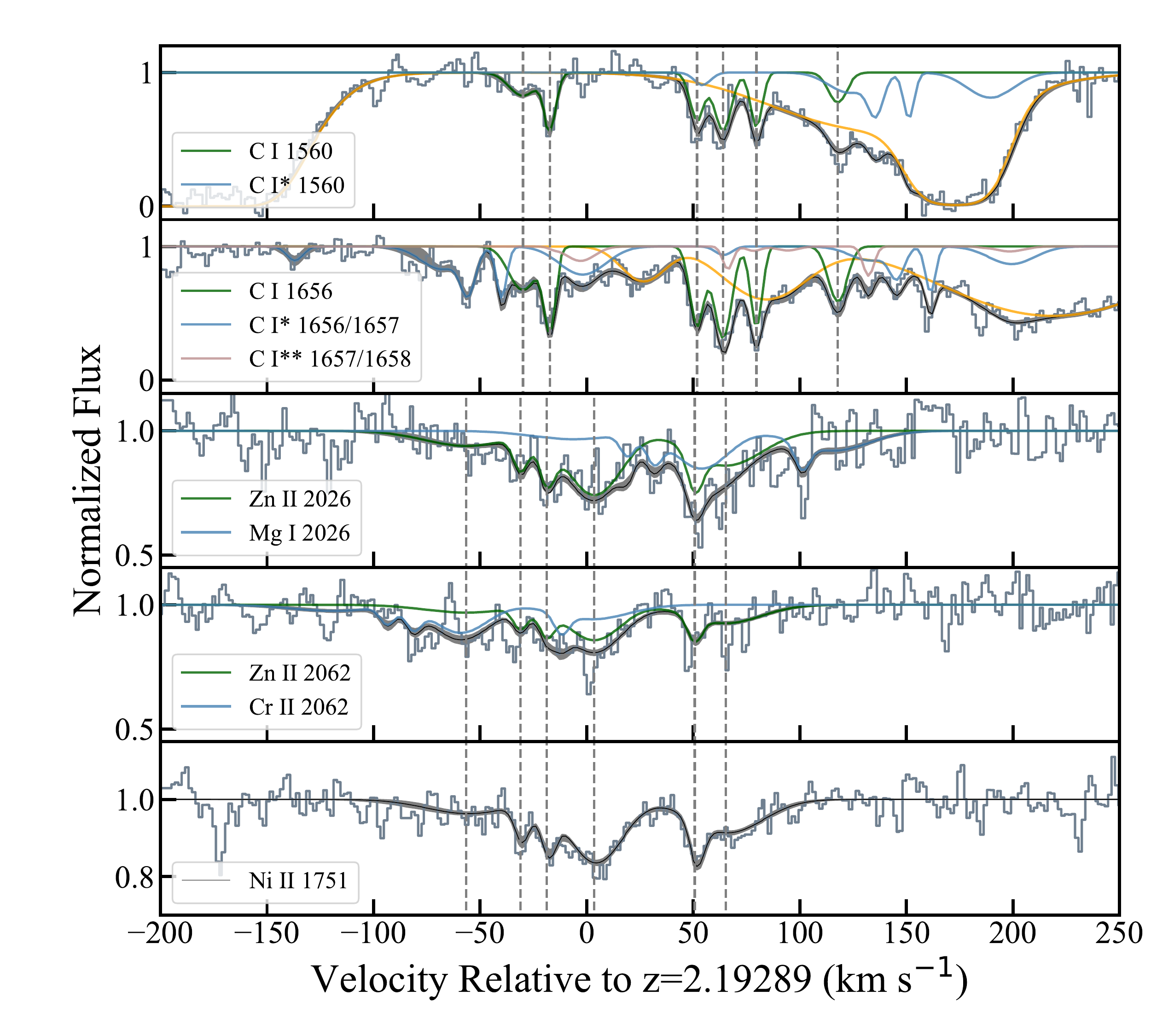}
\caption{Selection of absorption lines in the VLT-UVES spectrum. The median model (black line) and
1$\sigma$ uncertainties (gray region) from the MCMC fitting routine as well as individual metal lines are shown.
The vertical (gray) dashed lines mark the six velocity components of the C\,\textsc{i}, Zn\,\textsc{ii}, and 
Ni\,\textsc{ii} metal lines. The yellow lines in the top two panels are absorption features due to 
higher redshift intervening \HI\ systems. The velocity offsets for Mg\,\textsc{i} and 
Cr\,\textsc{ii} are 50~\kms\ and $-63$~\kms, respectively. Note the lack of C\,\textsc{i} absorption in the 
strongest Zn\,\textsc{ii} and Ni\,\textsc{ii} component at $v \approx 3 \; \rm km~s^{-1}$.}
\label{fig:Abs}
\end{figure}

\section{Observations}
\label{sec:obs}

\subsection{UVES Observations}
\label{sec:obs-UVES}

To search for metal lines from the DLA, a high resolution spectrum was obtained of 
QSO~B1228$-$113 using the Ultraviolet and Visual Echelle Spectrograph \citep[UVES;][]{Dekker2000} 
on the Very Large Telescope (VLT). Details of the observations and reduction procedures are described in
\citet{Akerman2005}. By fitting a single component to the Zn\,\textsc{ii} absorption line, these authors 
reported a metallicity of ${\rm [M/H]} = -0.22$. To provide an updated measure of the metallicity, we 
have renormalized the spectrum, and refitted the metal lines using a Monte Carlo Markov Chain Voigt profile 
fitting routine publicly available in the \textsc{linetools} \footnote{https://github.com/linetools/linetools}
package \citep{linetools}.

\begin{table}[!b]
\centering
\caption{Properties of the Absorber}
\label{tab:absprop}
\begin{tabular}{ll}
\multicolumn{2}{c}{DLA~B1228$-$113}\\
\hline
\hline
Right Ascension (J2000) & 12:30:55.62 \\
Declination (J2000) & -11:39:09.9\\
Redshift & 2.19289\\
$\log(N$(H\,\textsc{i})/cm$^{-2})$ & $20.60 \pm 0.10$\\
$\rm{[M/H]}$ & $-0.27 \pm 0.10$\\
$\Delta V_{90}$ (\kms) & $163 \pm 10$\\
$\log(N$(Zn\,\textsc{ii})/cm$^{-2})$ & $12.96 \pm 0.03$\\
$\log(N$(Mg\,\textsc{i})/cm$^{-2})$& $13.34 \pm 0.05$\\
$\log(N$(Cr\,\textsc{ii})/cm$^{-2})$ & $13.23 \pm 0.06$\\
$\log(N$(Ni\,\textsc{ii})/cm$^{-2})$& $14.05 \pm 0.02$\\
$\log(N$(C\,\textsc{i})/cm$^{-2})$ & $13.88^{+0.08}_{-0.04}$\\
$\log(N$(C\,\textsc{i}$^*$)/cm$^{-2})$ & $14.0 \pm 0.2$\\
$\log(N$(C\,\textsc{i}$^{**}$)/cm$^{-2})$ & $ < 13.5 (2\sigma)$\\
\HI\ 21\,cm Optical depth & $ < 0.13 (3\sigma)$\\
Covering factor, $f$ & 0.93\\
Spin temperature (K)  & $< 1895 \times (f/0.93)$\\
\hline
\end{tabular}
\end{table}

Specifically, we fit the Ni\,\textsc{ii} $\lambda 1741$ and Ni\,\textsc{ii} $\lambda 1751$
lines with six absorption components to provide an accurate model of the absorption profile.
We then tie the component structure of the blended lines (i.e., Zn\,\textsc{ii}, Mg\,\textsc{i}, and Cr\,\textsc{ii})
to the Ni\,\textsc{ii} lines assuming turbulent broadening \citep[see e.g.,][]{Prochaska1997}
leaving only the total column density as a variable. This yields a Zn\,\textsc{ii} column density of 
$\log(N$(Zn\,\textsc{ii})/cm$^{-2})$ $=12.96 \pm 0.03$, resulting in an updated metallicity 
of ${\rm [M/H]} = -0.27 \pm 0.10$, consistent with the previous measurement. The VLT-UVES spectrum
also showed absorption from the ground and excited fine structure states of neutral 
carbon (C\,\textsc{i}, C\,\textsc{i}$^*$ and  C\,\textsc{i}$^{**}$). A 6-component fit of this complex is 
shown in Figure \ref{fig:Abs}. The redshift, relative strength and Doppler parameters of the C\,\textsc{i} 
lines are different from the other low-ionization lines, as this line traces the coldest components of the 
gas \citep[e.g.,][]{Jorgenson2010}. The total column densities of the different species are listed 
in Table \ref{tab:absprop}.

\subsection{ALMA Observations}
\label{sec:obs-ALMA}

\begin{figure*}[t!]
\includegraphics[width=0.5\textwidth]{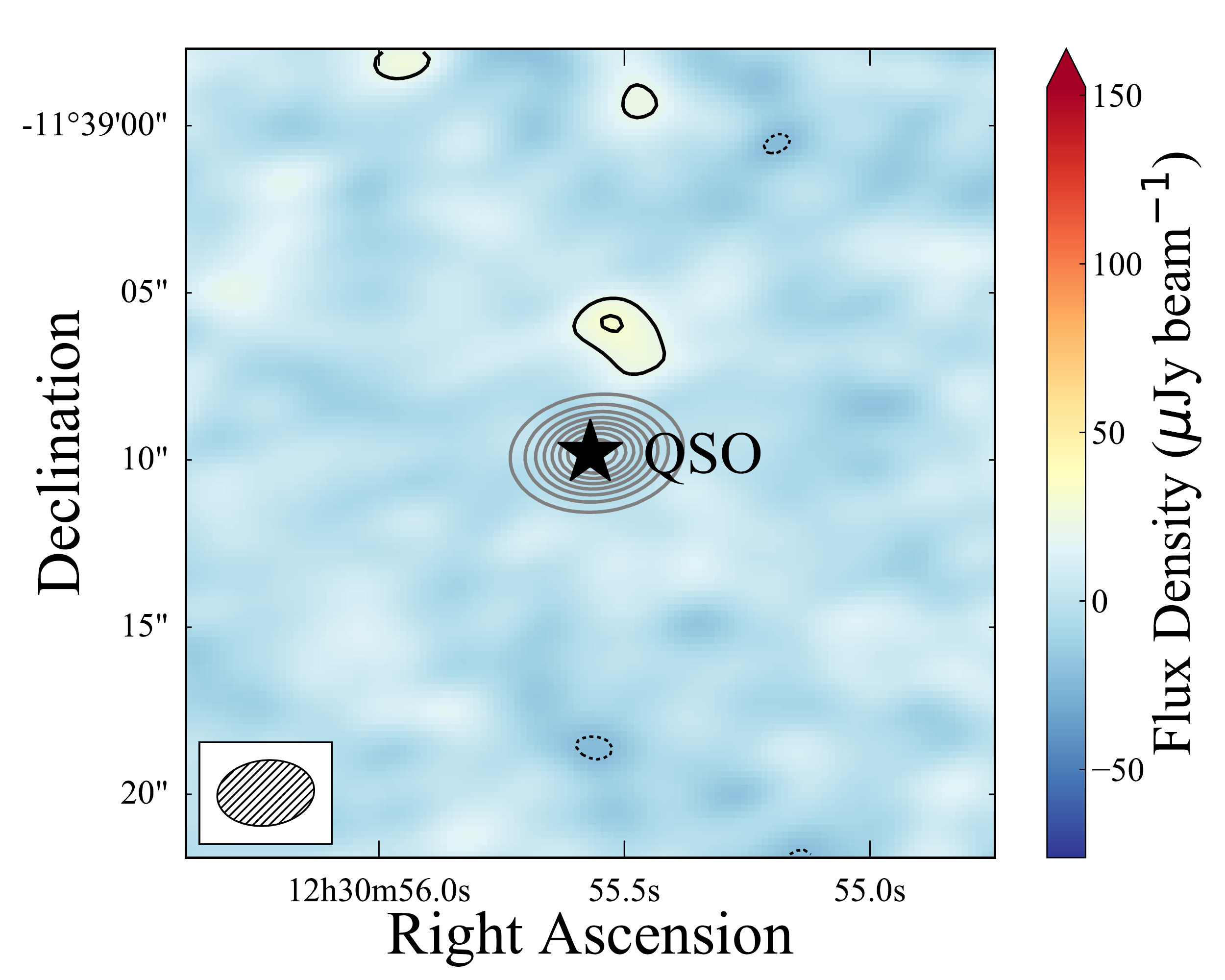}
\includegraphics[width=0.5\textwidth]{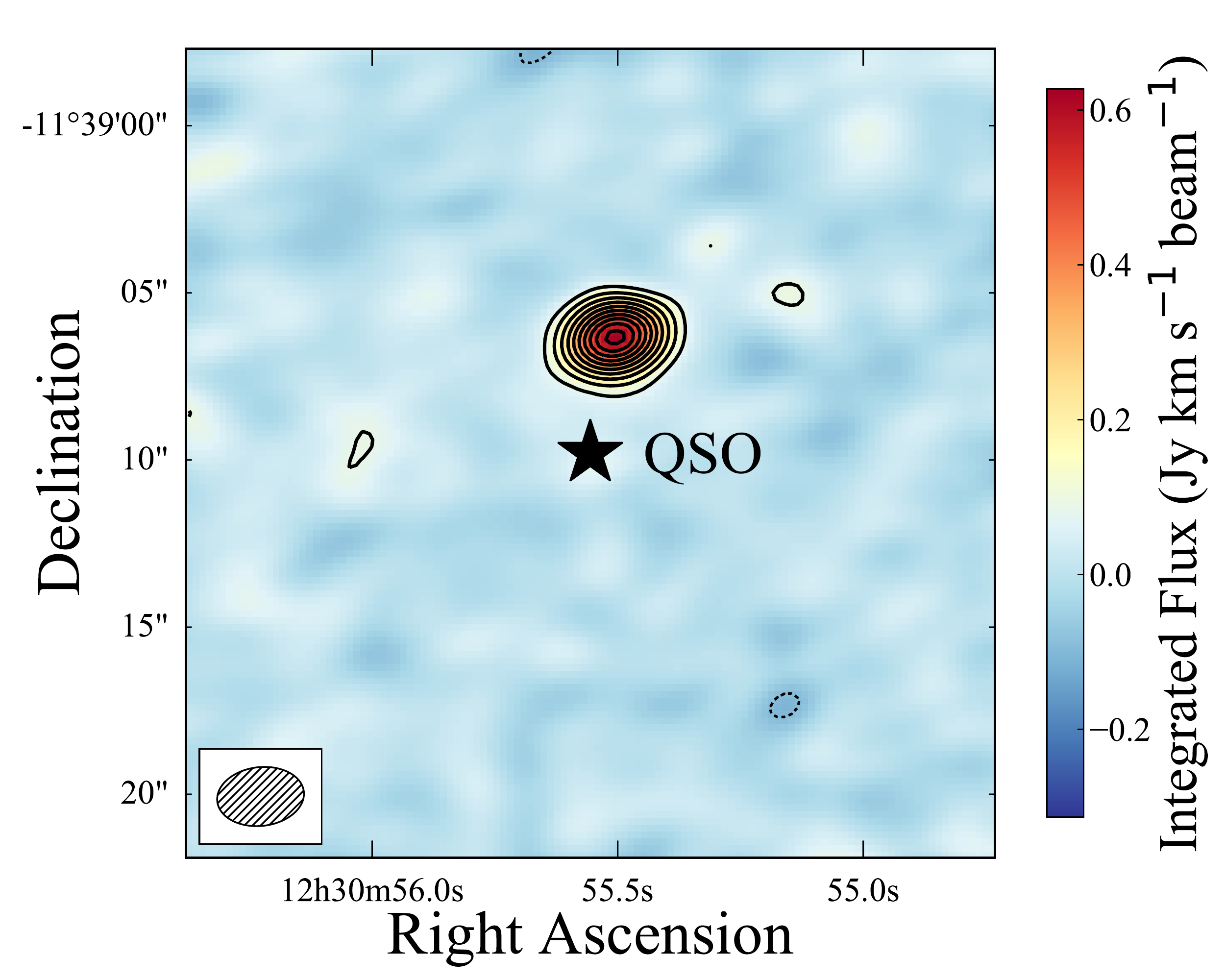}
\caption{Left: 100.5\,GHz continuum image of the field surrounding QSO~B1228$-$113. To highlight
the continuum emission of fainter sources, we have subtracted out the QSO emission (which was 
detected at very high signal-to-noise ratio). Gray contours start at 300$\sigma$ and increase by 300$\sigma$, 
whereas black contours start at 3$\sigma$ and increase by $\sqrt{2}\sigma$. Right: Integrated CO(3$-$2) 
emission from the channels showing line emission (Figure \ref{fig:Spec}). Contours begin at 3$\sigma$ and 
increase by 2$\sigma$. Dotted contours indicate negative values. The synthesized beam is 
shown in the bottom left inset.}
\label{fig:Cont}
\end{figure*}

The field surrounding QSO~B1228$-$113 was observed with ALMA on UT~2017~April~7 and 8 with 
a compact configuration (maximum baseline of 453~m) for a total on-source integration time of 
2.4~hours. One of the four spectral windows was centered on the redshifted CO($3-2$) line at 
108.3~GHz, and the remaining three spectral windows were set up to measure continuum 
emission. Callisto and Ganymede were used as flux calibrators, while QSO~J1256$-$0547 
and QSO~J1216$-$1033 were used for bandpass and phase calibration, respectively. 

The initial calibration was carried out using the ALMA pipeline in the Common Astronomy Software 
Applications \citep[CASA;][]{McMullin2007} package. The quasar continuum flux density (18.8~mJy; 
left panel Figure \ref{fig:Cont}) was sufficient to perform self-calibration, which was done in the Astronomical 
Image Processing System \citep[AIPS;][]{Greisen2003} package. The continuum image was created in CASA 
using natural weighting, resulting in a synthesized beam of $2.9'' \times 2.0''$ at $-83.8$\textdegree\ and a 
root mean square (RMS) noise of 6.9~$\mu$Jy~beam$^{-1}$. The spectral cube was Hanning-smoothed to 
a velocity resolution of 43.2~\kms. The resulting, naturally weighted spectral cube has a mean synthesized beam of 
$2.6'' \times 1.8''$ at $-83.4$\textdegree, and an RMS of 0.15~mJy~beam$^{-1}$ per 43.2~\kms\ channel.

A clear emission line is detected in the continuum-subtracted spectral cube with a full width at half 
maximum (FWHM) of $600 \pm 60$~\kms\ and a velocity-integrated flux density of $0.73 \pm 0.06$~Jy~\kms\ 
(Table \ref{tab:emsprop} and Figure \ref{fig:Spec}). The emission is spatially offset from the DLA by 
$3.5''$ at a position angle of -14\textdegree. Weak continuum emission is also detected at this location, with
a flux density of $46 \pm 10$~$\mu$Jy (Figure \ref{fig:Cont}). Both line and continuum emission 
from this source are spatially unresolved in the present ALMA images.

\begin{table}[b!]
\centering
\caption{Properties of the Galaxy}
\label{tab:emsprop}
\begin{tabular}{ll}
\multicolumn{2}{c}{ALMA~J123055.50$-$113906.4} \\
\hline
\hline
Right Ascension (J2000) & 12:30:55.50 \\
Declination (J2000) & -11:39:06.4 \\
Redshift of CO(3$-$2) emission & $2.1933 \pm 0.0005$\\
Redshift of H$\alpha$/[N\,\textsc{ii}] emission & $2.1912 \pm 0.0007$\\
$S_{\rm cont}$ at $\nu_{\rm obs}$ = 100.5 GHz ($\mu$Jy) & $46 \pm 10$ \\
FWHM$_{\rm CO(3-2)}$ (\kms) & $600 \pm 60$ \\
$ \int S_{\rm CO(3-2)} dv$ (Jy \kms) & $0.73 \pm 0.06$ \\
$L_{\rm CO(3-2)}$ (L$_\odot)$ & $(2.5 \pm 0.2) \times 10^{7}$ \\
$L'_{\rm CO(3-2)}~({\rm K~km~s^{-1}~pc^2})$ & $(1.88 \pm 0.15) \times 10^{10}$ \\
$f$(H$\alpha$) (erg s$^{-1}$ cm$^{-2}$) & $(2.0 \pm 0.2) \times 10^{-17}$ \\
$f$([N\,\textsc{ii}]) (erg s$^{-1}$ cm$^{-2}$) & $(0.5 \pm 0.3) \times 10^{-17}$ \\
$L_{\rm TIR}$ (L$_\odot)$ & $(2.2 \pm 0.5) \times 10^{12}$\\
M$_{\rm mol}$ (M$_\odot$) & $(1.4 \pm 0.2) \times 10^{11}$\\
SFR$_{\rm H\alpha}$ (M$_\odot$ yr$^{-1}$) & $3.9 \pm 0.4$\\
SFR$_{\rm dust-corrected}$  (M$_\odot$ yr$^{-1}$) & $110^{+60}_{-30}$\\
\hline
\end{tabular}
\end{table}

\subsection{SINFONI Observations}
\label{sec:obs-SINFONI}
The SINFONI spectrograph \citep{Eisenhauer2003} on the VLT was used to obtain near-infrared 
integral field spectroscopy of the field surrounding QSO~B1228$-$113 in program ID:~080.A-0742(A) 
\citep[PI Peroux; see][]{Peroux2011}. We re-reduced the data using the ESO SINFONI pipeline v.~2.9. 
The final co-added data cube was scaled to match the previously measured flux of the integrated 
QSO spectrum \citep{Ellison2005}. This was necessary because the response functions of the 
flux calibrators for each night showed variations of $\approx 50$\%.

\begin{figure*}[t!]
\hspace{-0.9cm}
\includegraphics[width=0.55\textwidth]{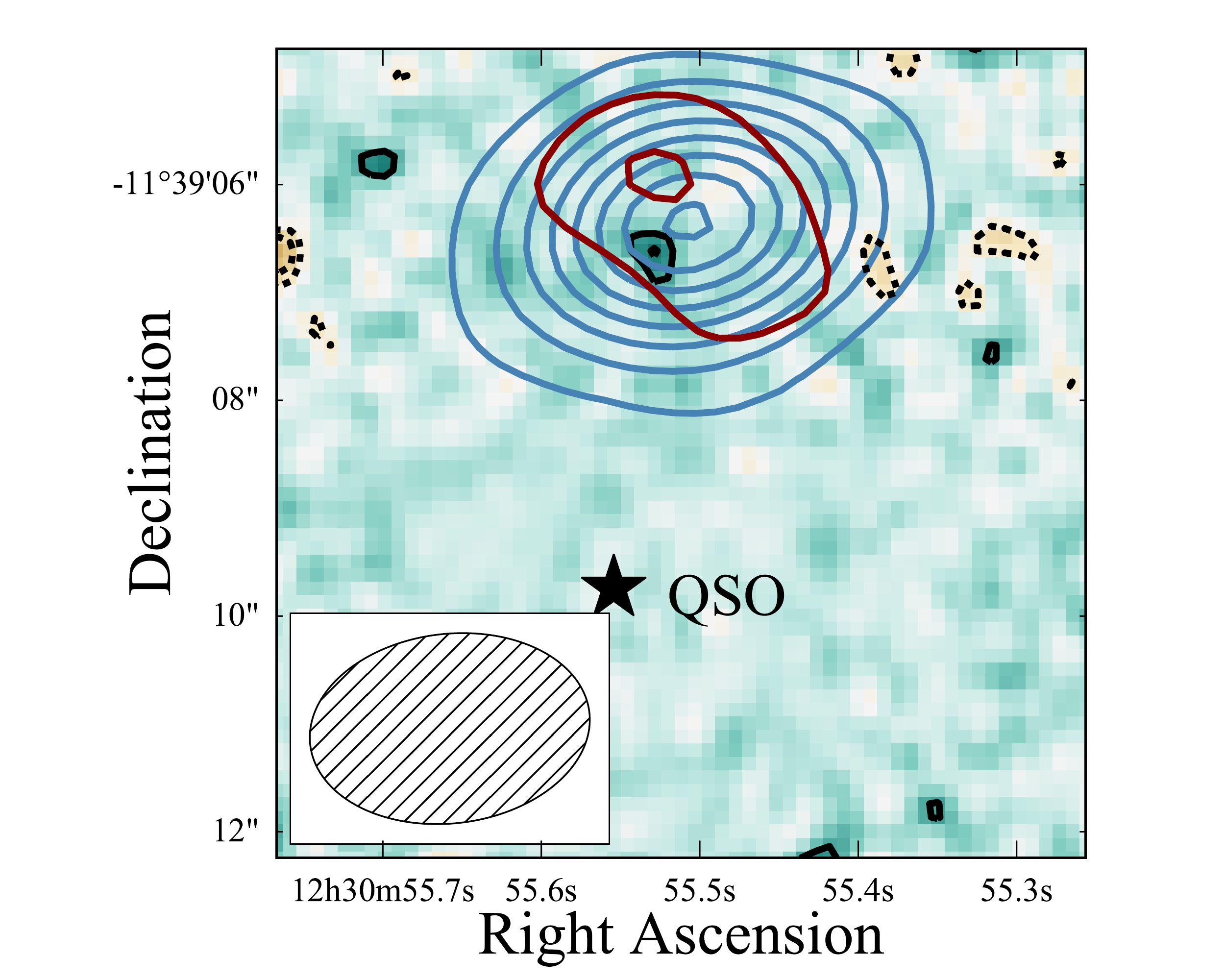}
\includegraphics[width=0.49\textwidth]{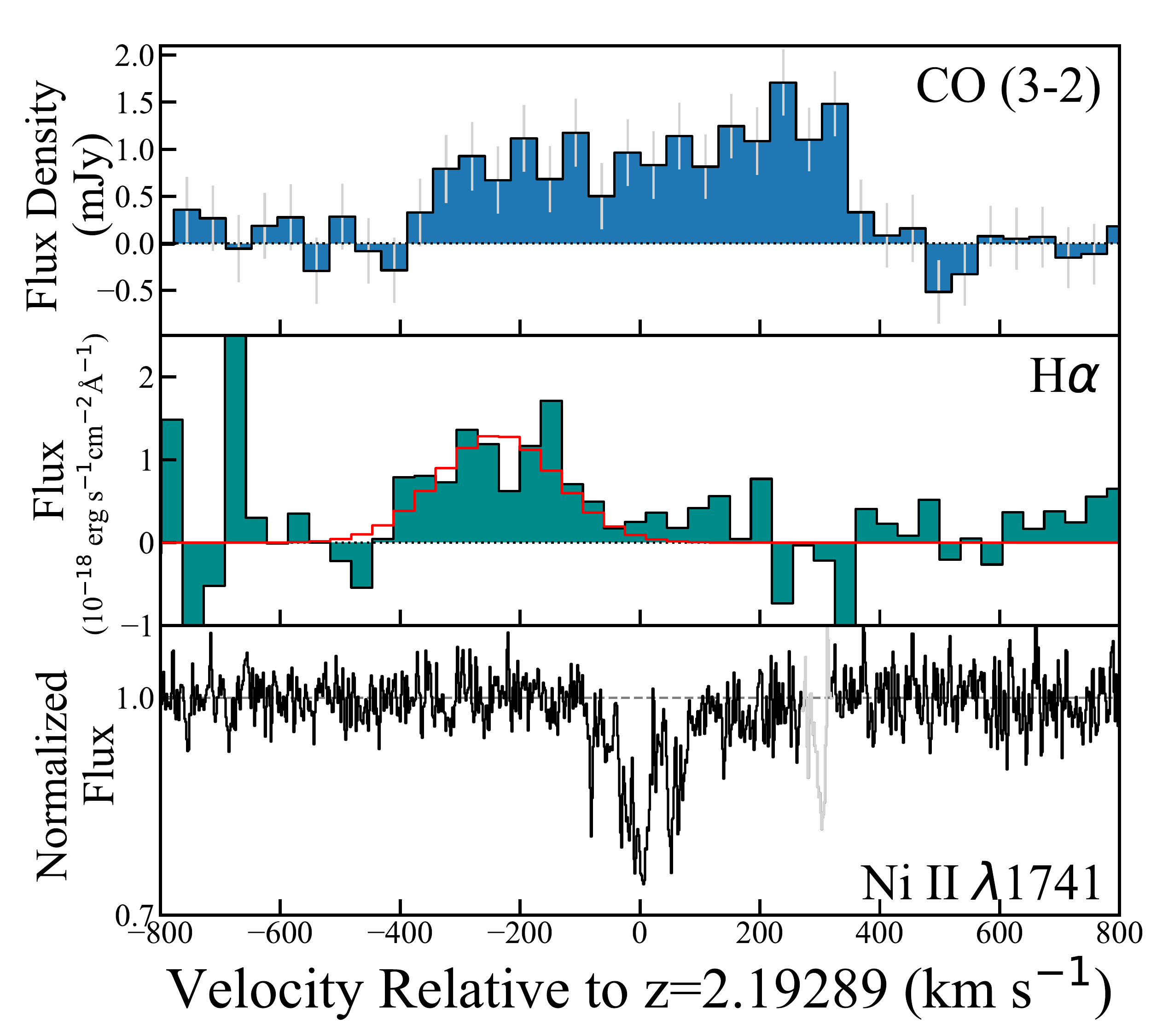}
\caption{Left: H$\alpha$ (image and black contours), CO (blue contours) and 100.5~GHz continuum 
(dark red contours) emission. Outer contours are at 3$\sigma$,  with the contours increasing 
by $\sqrt{2}\sigma$ for H$\alpha$ and the 100.5~GHz continuum, and by $3\sigma$ for the CO(3$-$2) 
line. The emission peaks of all lines lies well within the ALMA synthesized beam (bottom left inset). 
Right: Spectra of the CO and H$\alpha$ emission lines, and a reference absorption line from the DLA
(Ni\,\textsc{ii}~$\lambda$1741); the CO emission and DLA absorption occur at a similar redshift, whereas the
H$\alpha$ emission is slightly offset by $\sim 220$~\kms. This is similar to the absorber-galaxy pair discussed in 
\citet{Moller2018}, who speculated this could be due to varying absorption across the galaxy.
The Gaussian fit to the H$\alpha$ emission is shown in red.}
\label{fig:Spec}
\end{figure*}

To measure possible H$\alpha$ emission at the position of the ALMA CO(3$-$2) emission, we created a pseudo 
narrow-band image centered on the redshifted H$\alpha$ line at 2095.5~nm from the 
SINFONI data cube. Any potential continuum emission was subtracted from this image by interpolating the flux in 
adjacent wavelength intervals blueward and redward of the H$\alpha$ line. The one dimensional 
spectrum (Figure \ref{fig:Spec}: middle panel) was extracted with an aperture similar in size to the
FWHM seeing of the observation. The spectrum shows an emission line that we identify as 
H$\alpha$ at $z=2.1912$. Besides H$\alpha$, [N\,\textsc{ii}] $\lambda$6586 is marginally detected
as well.
We fitted both emission lines with the Image Reduction and Analysis Facility package \citep[IRAF;][]{Tody1993}
using \textsc{ngaussfit} and derived a total emission line flux of 
$f(\text{H}\alpha) = (2.0 \pm 0.2) \times 10^{-17}~\text{erg~s}^{-1}~\text{cm}^{-2}$ and
$f\text{([N\,\textsc{ii}])} = (0.5 \pm 0.3) \times 10^{-17}~\text{erg~s}^{-1}~\text{cm}^{-2}$, 
consistent with the previous flux limits \citep{Peroux2011}.

\subsection{GMRT Observations}
\label{sec:obs-GMRT}
The $250-500$~MHz receivers of the Giant Metrewave Radio Telescope (GMRT) were used to carry out 
a search for redshifted \HI~21\,cm absorption from the DLA towards QSO~B1228$-$113 on 2017 May~8.
The GMRT Software Backend was used as the correlator with a bandwidth of 2.08 MHz centered at 
444.85~MHz. The total velocity coverage is $\approx$1400~\kms\ at a velocity resolution of 
$2.7$~\kms. The total on-source time was $\sim$5~hours, with 25 working antennas.

The GMRT data were analyzed in AIPS using standard procedures for low-frequency imaging and 
spectroscopy \citep[e.g.,][]{Kanekar2014}. The final spectral cube has an RMS noise of 1.7~mJy 
per 2.7~km~s$^{-1}$ channel, while the measured quasar continuum flux density is
$340.4 \pm 0.6$~mJy. The quasar spectrum shows no evidence for \HI~21\,cm absorption, 
yielding a $3\sigma$ upper limit on the velocity-integrated \HI~21\,cm optical depth of 
$0.13$~km~s$^{-1}$, assuming a Gaussian line profile with a FWHM of 20~km~s$^{-1}$.
This implies a $3\sigma$ lower limit of ($2038 \times f$)~K to the DLA spin temperature, where $f$
is the DLA covering factor. \citet{Kanekar2009} used the Very Long Baseline Array (VLBA) to measure
the core flux density of the quasar; combining the VLBA 327~MHz core flux density with the new GMRT 
total flux density yields a DLA covering factor of $f=0.93$. The $3\sigma$ lower limit to the 
DLA spin temperature is then $T_{\rm s} > 1895 \times (f/0.93)$~K.
 
\section{Results}
\label{sec:results}

\subsection{Galaxy Properties}
We identify the line detected in the ALMA observations as the redshifted CO(3$-$2) emission line
at $z = 2.1933$, in excellent agreement with the DLA absorption redshift, $z = 2.19289$. 
The velocity integrated CO(3$-$2) flux density of $0.73 \pm 0.06$~Jy~\kms\ implies a
CO(3$-$2) line luminosity of $L'_{\text{CO(3$-$2)}}$ = $(1.88 \pm 0.15) \times 10^{10}$~K~\kms~pc$^2$ or 
$L_{\text{CO(3$-$2)}}$ = $(2.5 \pm 0.2) \times 10^{7}$~L$_\odot$. 

To estimate the total infrared luminosity[($L\text{(TIR)}$; defined as the integrated luminosity over 
the 8 $-$ 1000 $\mu$m wavelength range] from the ALMA 100.5~GHz continuum image, we
fitted a modified black-body spectrum with mid-infrared slope,  $\alpha = 1.5$, spectral emissivity 
index, $\beta = 1.5$, and dust temperature,  $\text{T}_{\text{dust}} = 35$\,K, to the measured flux 
density \citep[see e.g.,][]{Neeleman2017}. This yields a total infrared luminosity of 
 $L\text{(TIR)} = (2.2 \pm 0.5) \times 10^{12}~L_\odot$. This estimate has a large systematic 
 uncertainty ($\sim$0.6~dex) because both $\beta$ and 
$\text{T}_{\text{dust}}$ are not constrained by the single dust continuum measurement. We note, however, that
this estimate is consistent with the estimate obtained from our $L'\text{(CO)}$ measurement, using the 
relationship between $L'\text{(CO)}$ and $L\text{(TIR)}$ in high-$z$ galaxies 
\citep[$L\text{(TIR)} = (2.2 \pm 1.5) \times 10^{12}~\text{L}_\odot$;  see e.g.,][]{Carilli2013,Dessauges2015}.

From the H$\alpha$ detection, we estimate a dust-uncorrected star formation rate (SFR) of
$3.9 \pm 0.4$~M$_\odot$~yr$^{-1}$ \citep[assuming a Kroupa initial mass function;][]{Kennicutt2012}. 
Correcting this for dust obscuration using the total infrared luminosity \citep{Kennicutt2012} yields a dust-corrected 
SFR of $110^{+60}_{-30}$~M$_\odot$~yr$^{-1}$, where the uncertainties include the systematic uncertainty
on the total infrared luminosity. Comparison of the two SFRs suggest that the galaxy is highly dust-obscured, 
far more than typical galaxies at these redshifts \citep[e.g.,][]{Moustakas2006}. Only models with uncommonly 
low dust temperatures, $\text{T}_{\text{dust}} \lesssim 25~\text{K}$, yield more typical dust-obscuration values, 
with dust-corrected SFRs $\lesssim 20$~M$_\odot$~yr$^{-1}$.

The molecular mass of the galaxy is estimated from the CO(3$-$2) line luminosity assuming
a CO(3$-$2) to CO(1$-$0) line ratio of $r_{31} = L'_{\text{CO(3$-$2)}}/L'_{\text{CO(1$-$0)}} = 0.57$
\citep{Dessauges2015} and a CO-to-H$_2$ conversion factor of 
$\alpha_{\text{CO}} = 4.3~\text{M}_\odot~(\text{K~km~s}^{-1}~\text{pc}^2)^{-1}$. These assumptions
are valid for typical star-forming galaxies \citep{Bolatto2013}, and yield a total molecular gas mass estimate of 
$M_{\text{mol}} = (1.4 \pm 0.2) \times 10^{11} \times (\alpha_{\text{CO}}/4.3) \times (0.57/r_{31})$~M$_\odot$. 
This is at the upper end of the molecular gas mass distribution for star-forming galaxies at this redshift
\citep{Tacconi2013,Genzel2015}.

Note that if physical conditions are more akin to those in starburst galaxies, then 
$\alpha_{\text{CO}} \approx 1~\text{M}_\odot~(\text{K~km~s}^{-1}~\text{pc}^2)^{-1}$ and $r_{31} \approx 1$,
yielding a molecular gas mass lower by a factor of $\approx 10$. However, we disfavor this scenario,
as even the dust-corrected SFR estimate is significantly below the median SFRs seen in starburst
galaxies \citep[e.g.][]{Daddi2005}. We emphasize that, despite the uncertainty in $\alpha_{\rm CO}$ and $r_{31}$,
this galaxy is likely massive, an assertion that is corroborated by the large CO(3$-$2) line 
width, FWHM~$\approx 600 \pm 60$~\kms \citep{Tiley2016}.

\subsection{Galaxy/DLA Association}
The agreement in redshift between the galaxy and DLA, as well as the low angular separation of the galaxy 
from the quasar sightline ($\approx 3.5''$ or $\approx 30$~kpc at the DLA redshift),
confirms the association between the molecular gas-rich galaxy and the high-metallicity absorber. 
This is consistent with results found at lower redshift, where CO emission studies of high-metallicity 
DLAs at $z \approx 0.7$ \citep{Moller2018,KanekarPrep} suggest that the cross-section for such DLAs appears 
to be biased toward galaxies with high molecular masses. Specifically, for 5 out of 7 targeted intermediate-redshift,
high-metallicity absorbers, a galaxy with high molecular gas mass was found within $\approx 50$~kpc of the 
DLA, and at the absorber redshift \citep{KanekarPrep}. Clearly a larger sample of CO observations surrounding 
high-metallicity $z \sim 2$ DLAs is needed to confirm this assertion at these redshifts.

Of course there always is a possibility that the Ly-$\alpha$ absorption originates from a lower mass galaxy below 
the sensitivity threshold of the present ALMA observations. However, the VLT-SINFONI observations yield an 
upper limit to the SFR of such a putative galaxy of $< 0.9$~M$_\odot$~yr$^{-1}$ \citep{Peroux2011}.
In addition, we note that the excellent agreement in velocity between the centroid of the CO emission and 
the low-ionization metal line absorption ($\approx 40$~\kms; see Figure~\ref{fig:Spec}) disfavors scenarios 
where the absorber is probing gas with a bulk motion with respect to the galaxy --- e.g., a single satellite galaxy 
or an extended disk --- as in this case one would expect to see a net velocity offset between 
absorption and emission \citep{Neeleman2016b}. The large velocity width of the DLA, 
$\Delta V_{90} = 163 \pm 10$~\kms \citep[see e.g.,][]{Prochaska1997} is further evidence the DLA 
is associated with a more massive galaxy halo, \citep[e.g.,][]{Ledoux2006} and large stellar mass 
\citep[$\log(M_{*}/{\rm M}_\odot) \gtrsim10.5$;][]{Christensen2014}.

\subsection{Physical Conditions of the Absorbing Gas}
The detection of C\,\textsc{i} in the UVES spectrum, indicates the presence of cold dense gas in the DLA, as
C\,\textsc{i} has been linked to the presence of molecular hydrogen \citep[e.g.,][]{Srianand2005}. 
The C\,\textsc{i} absorption in the $z = 2.19289$ DLA spans a wide range of velocities, indicating 
that the line-of-sight probes several distinct cold gas clumps. Interestingly, the strongest 
absorption component of the dominant low-ionization lines (e.g., Zn\,\textsc{ii}) at $v = 3$~\kms\ shows 
no C\,\textsc{i} absorption. This suggests that this component must be significantly warmer and less dense 
than the C\,\textsc{i}-bearing components.

This scenario is corroborated by our \HI~21\,cm absorption measurement. The high spin temperature,
($T_{\rm s} > 1895 \times (f/0.93)$~K), is inconsistent with the known anti-correlation between $T_{\rm s}$ 
and metallicity [M/H] \citep{Kanekar2014}. If, however, we assume that the \HI\ is predominantly associated with 
the strongest Zn\,\textsc{ii} absorption component, the metallicity of this gas is reduced to $-0.7$, which is 
consistent with the $T_{\rm s}$--[M/H] anti-correlation. We note that this requires that approximately 
50\,\% of the metals are locked up in the denser phase traced by C\,\textsc{i}, and that this phase contribute little 
to the total \HI\ column density. 

This implies that either this phase is very metal-rich and small, thereby containing intrinsically little \HI\ gas, or 
that most of the \HI\ has been converted into H$_2$. However, a rough estimate of the H$_2$ column density 
from the expected scaling relations between C\,\textsc{i} and H$_2$ \citep{Glover2016} gives an H$_2$ column 
density of $4 \times 10^{18}$~cm$^{-2}$, well below the total H\,\textsc{i} content of the DLA but consistent with
previous molecular hydrogen column density measurements in DLAs \citep[e.g.,][]{Srianand2005}. We therefore 
favor the first scenario whereby most of the metals are locked up in small metal-rich clumps. Similar 
multi-phase structure has been previously observed in a few high-$z$ absorbers 
\citep[e.g.,][]{Noterdaeme2017, Rudie2017}. Alternatively, the high inferred spin temperature might arise if 
the sightline towards the radio core has a significantly lower \HI\ column density than that towards the 
optical QSO \citep[e.g.,][]{Wolfe2003,Kanekar2014}.

\section{Summary}

We present, for the first time, molecular emission from a galaxy associated with a DLA at $z \sim 2.2$. Our results
highlight the ability of ALMA to detect and characterize the galaxies associated with high-metallicity DLAs 
at the peak epoch of galaxy assembly. We obtain a high molecular gas mass, 
$M_{\text{mol}} = (1.4 \pm 0.2) \times 10^{11} \times (\alpha_{\text{CO}}/4.3) \times (0.57/r_{31})$~M$_\odot$,
at the upper end of the mass distribution for star-forming galaxies at these redshifts \citep[e.g.,][]{Tacconi2013,Genzel2015}.
The detection of far-infrared continuum with ALMA and weak H$\alpha$ emission indicates significant 
amounts of dust obscuration, and a dust-corrected SFR of $\approx 110^{+60}_{-30}$~M$_\odot$~yr$^{-1}$.

The high molecular gas mass and large impact parameter ($\approx 30$~kpc) are consistent with a scenario in
which high-metallicity DLAs typically arise in the near-vicinity of massive gas-rich galaxies. Finally,
the detection and non-detection, respectively, of C\,\textsc{i} and \HI~21\,cm absorption suggest that the \HI\ 
along the sightline is predominantly warm, but that there are several cold dense gas components that contain
$\approx 50$\% of the metals. The $z=2.19289$ DLA towards QSO~B1228$-$113 thus highlights the power of 
combining absorption spectroscopy with emission line studies in order to study the multi-phase structure 
of the gas surrounding high redshift galaxies.

\acknowledgements
This paper makes use of ALMA data: ADS/JAO.ALMA\#2016.1.00628.S. ALMA is a partnership 
of ESO (representing its member states), NSF (USA) and NINS (Japan), together with NRC (Canada) and 
NSC and ASIAA (Taiwan) and KASI (Republic of Korea), in cooperation with the Republic of Chile. The 
Joint ALMA Observatory is operated by ESO, AUI/NRAO and NAOJ. The National Radio Astronomy Observatory 
is a facility of the National Science Foundation operated under cooperative agreement by Associated 
Universities, Inc. MN acknowledges support from ERC Advanced Grant 740246 (Cosmic\_Gas). 
NK acknowledges support from the Department of Science and Technology via a Swarnajayanti Fellowship 
(DST/SJF/PSA-01/2012-13). LC acknowledges support from DFF$-$4090-00079. 
The Cosmic Dawn Center is funded by the DNRF.

\end{document}